\documentclass[doublecol]{epl2} 

\usepackage[T1]{fontenc}
\usepackage[english]{babel}
\usepackage{amsmath}
\usepackage{amssymb}
\usepackage{graphicx}
\usepackage[unicode=true,pdfusetitle,
 bookmarks=true,bookmarksnumbered=false,bookmarksopen=false,
 breaklinks=false,pdfborder={0 0 1},backref=false,colorlinks=false]
 {hyperref}

\usepackage{xcolor}
\usepackage{array}
\usepackage{multirow}

\usepackage[switch]{lineno}

\providecommand{\tabularnewline}{\\}

\title{Tunable phononic transparency and opacity with isotopic defects}
\author{Zhun-Yong Ong \thanks{E-mail: \email{ongzy@a-star.edu.sg} }  }

\institute{                    
  Institute of High Performance Computing (IHPC), Agency for Science,
  Technology and Research (A{*}STAR), 1 Fusionopolis Way, \#16-16 Connexis,
  Singapore 138632, Republic of Singapore
}

\abstract{
In an isotopically disordered harmonic chain, phonon transmission
attenuates exponentially with distance because of multiple scattering
by the isotopic defects. We propose a simple and systematic method,
which is based on the static structure factor, for arranging the isotopic
defects to suppress or enhance phonon scattering within a targeted
frequency window (TFW), resulting in frequency-selective maximization
or minimization of phonon transmission. The phononic transparency
and opacity effects are demonstrated numerically from the changes
in the transmission coefficient and the phonon inverse particpation
ratio. We briefly discuss how the underlying concept can be extended
to the design of aperiodic superlattices to improve or block phononic
transmission and control coherent phonon transport. 
}

\begin{document}

\maketitle

\section{Introduction}
Phonon scattering by defects in
a monoatomic harmonic chain, also known as a linear spring-mass system,
continues to be a subject of considerable interest, because of its
importance for understanding heat conduction in solids.~\cite{RJRubin:JMP68_Transmission,RJRubin:JMP70_Transmission2,RJRubin:JMP71_Abnormal,WLGreer:JMP72_Quantum,ADhar:PRL01_DisorderedHarmonicChain,ADhar:AdvPhys08_Heat,ZYOng:PRB14_Enhancement,YHu:NL25_Exactly}
A particular model of interest is the one-dimensional (1-D) lattice
with \emph{configurational} isotopic disorder, in which a host atom
is replaced by an isotopic defect (ID) at random lattice sites. In
this model, the interatomic force constants are identical but the
atomic mass is dichotomic with a value of either $m$ for the host
atom or $m_{\text{id}}$ for the ID atom, unlike other mass disorder
models where each atomic mass is a continuous random variable. The
atomic mass difference between a host atom and the ID ($\Delta m=m_{\text{id}}-m$)
leads to scattering when the phonon interacts with the ID.~\cite{PAllen:PRB22_Heat,ZYOng:PRB24_Exact}
In addition to its relevance for understanding phonon-mediated heat
conduction in real materials, the form of disorder is also important
for understanding wave transmission through aperiodic 1-D superlattices.~\cite{SJu:PRX17_Designing,RHu:PRB21_Direct,WDing:NL24_OptimallySuppressed,ZYan:APL09_Analysis,MMiniaci:PRAppl23_Spectral} 

Izrailev and co-workers~\cite{FIzrailev:PRL99_Localization,FIzrailev:PR12_Anomalous}
have shown that specific long-range correlations in the disorder for
the continuous on-site potential in 1-D tight-binding models can lead
to well-defined mobility edges, with the inverse localization length
directly dependent on the power spectrum of the on-site potential.
{
Hence, by tuning the power spectrum of the disorder, one can enhance
localization or delocalization of the modes.~\cite{UKuhl:APL00_Experimental,UKuhl:PRL08_Enhancement,ODietz:NJP12_Transmission} }
This insight is extended
in Refs.~\cite{IFHerreraGonzalez:EPL10_Anomalous,IFHerreraGonzalez:PRE23_Localization} to classical harmonic
chains with spatially correlated amplitudinal disorder in which the
atomic mass is a \emph{continuous} variable. 

{
Despite this advance, it remains challenging to extend these insights
to models with \emph{discrete} configurational disorder, 
where the disorder takes on the form of aperiodically distributed scatterers that are identical. } 
In this work,
we use this insight from Ref.~\cite{FIzrailev:PRL99_Localization}
as the basis of our approach to tuning the phononic transparency through
a harmonic chain with configurational disorder. Unlike them, we do
not try to map the mass distribution to a pre-determined power spectrum.
Instead, we are only interested in maximizing or minimizing transmission
through a finite \emph{targeted frequency window} (TFW), by optimizing
the static structure factors for the ID's.

In our paper, we derive the relationship between the localization
length $\gamma$ and the static structure factor $S_{\text{id}}$
for the ID distribution. We describe how $S_{\text{id}}$ can be minimized
(maximized) to suppress (enhance) scattering and generate phononic
transparency (opacity) in a TFW, through a simulated annealing process
that optimizes the positions of the ID's. We then illustrate the method
by simulating the frequency-dependent phonon transmission and localization.
Finally, we discuss the potential applications of our approach.

\section{Model of disorder in 1-D lattice}
We consider
a disordered 1-D harmonic lattice of $N$ sites with masses $m_{n}$,
where $n$ denotes the site index and $1\leq n\leq N$, a lattice
constant of $a$ like in Ref.~\cite{SZakeri:PRE15_Localization},
and periodic boundary conditions. The equation of motion for the $n$-th
atom is~ 
\begin{equation}
m_{n}\frac{\partial^{2}u_{n}}{\partial t^{2}}=-\zeta(u_{n}-u_{n-1})-\zeta(u_{n}-u_{n+1})\ ,\label{eq:EquationOfMotion}
\end{equation}
where $u_{n}$ denotes the displacement of the $n$-th atom from its
equilibrium position at $x_{n}=na,$ and $\zeta$ denotes the near-neighbor
spring constant, which we set $\zeta=1$ for simplicity. 

The atomic mass is given by $m_{n}=(1-c_{n})m+c_{n}m_{\text{id}}$,
where $c_{n}$ is the site occupation factor. We have $c_{n}=1$ if
the site is occupied by an ID and $0$ otherwise. Hence, each instance
of the configurational disorder is described by the binary sequence
$(c_{1},c_{x},\ldots,c_{N})$. The total number of ID's is $N_{\text{id}}=\sum_{n=1}^{N}c_{n}$
and the ID concentration is given by $c=N_{\text{id}}/N=\overline{c_{n}}$,
where the overhead line denotes the \emph{site average}. The site-averaged
mass is $\overline{m_{n}}=m+c(m_{\text{id}}-m)=m+c\Delta m$ and the
mass variance is $\overline{\delta m_{n}^{2}}=c(1-c)\Delta m^{2}$.
Finally, we can associate a phonon dispersion with $\overline{m_{n}}$
, \emph{i.e.}, $\omega_{k}^{2}=\omega_{\text{max}}^{2}\sin^{2}(\frac{ka}{2})$
where $\omega_{\text{max}}^{2}=4\zeta/\overline{m_{n}}$. 

To describe how the ID distribution affects phonon transmission, we
relate the ID distribution to the mass distribution, especially in
terms of the spatial correlation of the mass. We characterize the
\emph{positional} distribution of the ID's by its static structure
factor 
\begin{equation}
S_{\text{id}}(q)=\frac{1}{N_{\text{id}}}\left|\sum_{p=1}^{N_{\text{id}}}e^{iqr_{p}}\right|^{2}\ ,\label{eq:StaticStructureFactor}
\end{equation}
where $r_{p}$ denotes the equilibrium position of the $p$-th ID
and $q=\frac{2\pi n}{Na}$ is the wave vector for $0\leq n<N$. We
note that for $q\neq0$, $S_{\text{id}}(q)$ is a function of the
positions $(r_{1},\ldots,r_{N_{\text{id}}})$ of the $N_{\text{id}}$
defects and thus can be tuned by varying $r_{1},\ldots,r_{N_{\text{id}}}$. 

To describe the \emph{mass} distribution, we define the normalized
\emph{binary correlator} for mass as 
\begin{equation}
\chi(n)=\frac{\overline{\delta m_{l}\delta m_{l+n}}}{\overline{\delta m_{l}\delta m_{l}}}=\frac{\overline{c_{l}c_{l+n}}-c^{2}}{c(1-c)}\ ,\label{eq:BinaryCorrelator}
\end{equation}
where $\delta m_{l}=m_{l}-\overline{m_{n}}$. Equation~(\ref{eq:BinaryCorrelator})
depends only on $n$ because the site-averaged disorder is spatially
invariant. Equation~(\ref{eq:BinaryCorrelator}) implies that $\chi(n)=\chi(-n)$
and yields $\chi(0)=1$, since $c_{l}c_{l}=c_{l}$, and $\chi(n)=0$
for $n\neq0$ if there is no correlation in the distribution of the
IDs. The long-range correlations in $\chi(n)$ are characterized by
the \emph{density power spectrum} $W(q)=\sum_{n=-\infty}^{\infty}\chi(n)e^{iqna}=1+2\sum_{n=1}^{\infty}\chi(n)\cos(qna)$
for $-\frac{\pi}{a}<q\leq\frac{\pi}{a}$ in the thermodynamic ($N\rightarrow\infty$)
limit. 

\section{Localization length}
In Ref.~\cite{IFHerreraGonzalez:EPL10_Anomalous},
it is shown using second-order perturbation theory and assuming weak
disorder, that the density power spectrum is proportional to the the
\emph{inverse localization length} $\gamma(k)$ via the relationship
\begin{equation}
\gamma(k)=\frac{\overline{\delta m_{n}^{2}}}{2\overline{m_{n}}^{2}}\frac{\omega_{k}^{2}}{\omega_{\text{max}}^{2}-\omega_{k}^{2}}W(2k)\label{eq:InvLocalLength}
\end{equation}

To determine how the ID distribution affects $\gamma(k),$ we relate
Eq.~(\ref{eq:StaticStructureFactor}) to the density power spectrum
$W(q)$ by rewriting $S_{\text{id}}(q)$ as 
\begin{equation}
S_{\text{id}}(q)=\frac{1}{cN}\left|\sum_{n=1}^{N}c_{n}e^{iqna}\right|^{2}\label{eq:StructureFactorConfigDisorder}
\end{equation}
which, in the $N\rightarrow\infty$ limit for $q\neq0$, simplifies
to $\lim_{N\rightarrow\infty}S_{\text{id}}(q)=(1-c)W(q)$ and yields
$1-c$ in the absence of correlation. This allows us to rewrite $\gamma(k)$
as
\begin{equation}
\gamma(k)=\frac{\mathcal{F}(c)\omega_{k}^{2}}{\omega_{\text{max}}^{2}-\omega_{k}^{2}}\lim_{N\rightarrow\infty}S_{\text{id}}(2k)\ ,\label{eq:InvLocalLengthPowerSpectrum}
\end{equation}
where $\mathcal{F}(c)=\frac{c\Delta m^{2}}{2(m+c\Delta m)^{2}}$.
Equations~(\ref{eq:StaticStructureFactor}), (\ref{eq:StructureFactorConfigDisorder})
and (\ref{eq:InvLocalLengthPowerSpectrum}) describe the dependence
of the inverse localization length on the the configurational disorder.
They also imply that we can modify $S_{\text{id}}(q)$ and hence the
transmission spectrum by changing the positions $(r_{1},\ldots,r_{N_{\text{id}}})$
of the isotopic defects, because the frequency-dependent transmittance
$\Xi(\omega)$ at $\omega=\omega_{k}$ in the thermodynamic limit
is given by $\lim_{N\rightarrow\infty}\Xi(\omega_{k})=\exp[-2\gamma(k)L]$
for $L=Na$. 

\section{Generation of simulated disordered structures}
Equation~(\ref{eq:InvLocalLengthPowerSpectrum}) implies that the
value of $S_{\text{id}}(2k)$ and hence $\gamma(k)$ at the frequency
$\omega_{k}$ can be optimized for scattering suppression or enhancement
by varying $r_{1},r_{2},\ldots,r_{N_{\text{id}}}$ or equivalently,
the $N$ occupation factors $c_{1},c_{x},\ldots,c_{N}$. If $\gamma(k)$
is minimized (maximized), then phonon transmission is enhanced (suppressed)
at $\omega=\omega_{k}$ . More generally, we are interested in suppressing
or enhancing scattering over a finite collection of frequency points
within a TFW between $\omega_{\text{lo}}$ and $\omega_{\text{hi}}$.
These frequency points are more conveniently represented by a finite
set of wave numbers, given by $k=2\pi n/(Na)$, where $0\leq n<N$
is an integer, and $k_{\text{lo}}\leq k<k_{\text{hi}}$, with $k_{\text{lo}}=\frac{2}{a}\arcsin(\omega_{\text{lo}}/\omega_{\text{max}})$
and $k_{\text{hi}}=\frac{2}{a}\arcsin(\omega_{\text{hi}}/\omega_{\text{max}})$.
We denote this set of wave numbers by the symbol $\mathcal{K}$. The
size of $\mathcal{K}$, which we denote by $M(\mathcal{K})$, is limited
by $N_{\text{id}}$. 

Given $\mathcal{K}$, we associate the scattering suppression or enhancement
with an objective function 
{
\begin{equation}
E_{\mathcal{K}}(c_{1},c_{x},\ldots,c_{N})= \alpha\mu_{\text{id}}+\sigma_{\text{id}}\ ,\label{eq:ObjectiveFunction}
\end{equation}
}
where $\mu_{\text{id}}=\frac{1}{M(\mathcal{K})}\sum_{k\in\mathcal{K}}S_{\text{id}}(2k)$
and $\sigma_{\text{id}}=\sqrt{\frac{1}{M(\mathcal{K})}\sum_{k\in\mathcal{K}}|S_{\text{id}}(2k)-\mu_{\text{id}}|^{2}}$
denote the mean and standard deviation of $S_{\text{id}}(2k)$, defined
in Eq.~(\ref{eq:StructureFactorConfigDisorder}), in $\mathcal{K}$,
respectively. The two terms on the righthand side of Eq.~(\ref{eq:ObjectiveFunction})
ensure that the $S_{\text{id}}(2k)$ values are minimized or maximized
evenly, as in Ref.~\cite{VRomera-Garcia:PRAppl19_Stealth}. In Eq.~(\ref{eq:ObjectiveFunction}),
we set $\alpha=1$ ($\alpha=-1$) to suppress (enhance) scattering.
To facilitate the minimization of $E_{\mathcal{K}}$, we limit $M(\mathcal{K})<N_{\text{id}}/2$.
{
  We note that Eq.~(\ref{eq:StaticStructureFactor}) implies the sum rule $\frac{1}{N} \sum_q S(q) = 1$. 
  This means that during the minimization process, 
  the reduction (gain) in the spectral weight of $S(q)$ within the TFW must be exactly offset by the overall gain (reduction) outside of the TFW.
  It also means that the transmission gain (reduction) within the TFW is accompanied by an average transmission reduction (gain) outside of the TFW.
}

To minimize $E_{\mathcal{K}}$, we perform simulated annealing (SA).
At each SA step, the values of two randomly picked occupation factors
$c_{i}$ (occupied) and $c_{j}$ (unoccupied) are swapped to create
a new disorder configuration. We compute the energy difference $\Delta E_{\mathcal{K}}$
between the new and old disorder configurations. If $\Delta E_{\mathcal{K}}<0$,
then the new disorder configuration is accepted; if $\Delta E_{\mathcal{K}}>0$
and $\exp(-\beta\Delta E_{\mathcal{K}})>R$, where $\beta$ denotes
the SA inverse temperature parameter and $R$ is a randomly sampled
number between 0 and 1, then the new disorder configuration is also
accepted. Otherwise, the new disorder configuration is rejected. We
find that for a system of $N=4000$ and $N_{\text{id}}=400$, $2\times10^{7}$
steps are sufficient to reach a steady-state minimum of Eq.~(\ref{eq:ObjectiveFunction})
for either scattering suppression or enhancement. 

For a specified $\mathcal{K}$ defined by $k_{\text{lo}}$ and $k_{\text{hi}}$,
different final steady-state configurations can be reached, depending
on the initial condition. Hence, we group them into an \emph{ensemble}
of disordered configurations for each $\mathcal{K}$ associated with
a unique pair of $k_{\text{lo}}$ and $k_{\text{hi}}$. 

To complement the static structure factor $S_{\text{id}}(2k)$ from
Eq.~(\ref{eq:StaticStructureFactor}) that characterizes the disorder
configuration, we also define the \emph{transmission-implied structure
factor} for $\omega\leq\omega_{\text{max}}$,
\begin{equation}
S_{\text{tr}}(\omega)=-\frac{\omega_{\text{max}}^{2}-\omega^{2}}{2L\mathcal{F}(c)\omega^{2}}\left\langle \log\Xi(\omega)\right\rangle \ ,\label{eq:EffectiveTransportStructureSpectrum}
\end{equation}
where $\langle\ldots\rangle$ denotes the ensemble average for configurations
with the same $\mathcal{K}$. Equation~(\ref{eq:EffectiveTransportStructureSpectrum})
measures how closely the approximation in Eq.~(\ref{eq:InvLocalLengthPowerSpectrum}),
which is derived for the thermodynamic limit, predicts the attenuation
in phonon transmission for a finite system.

\section{Simulation setup} 
\begin{figure}
\begin{centering}
\includegraphics[width=8.6cm]{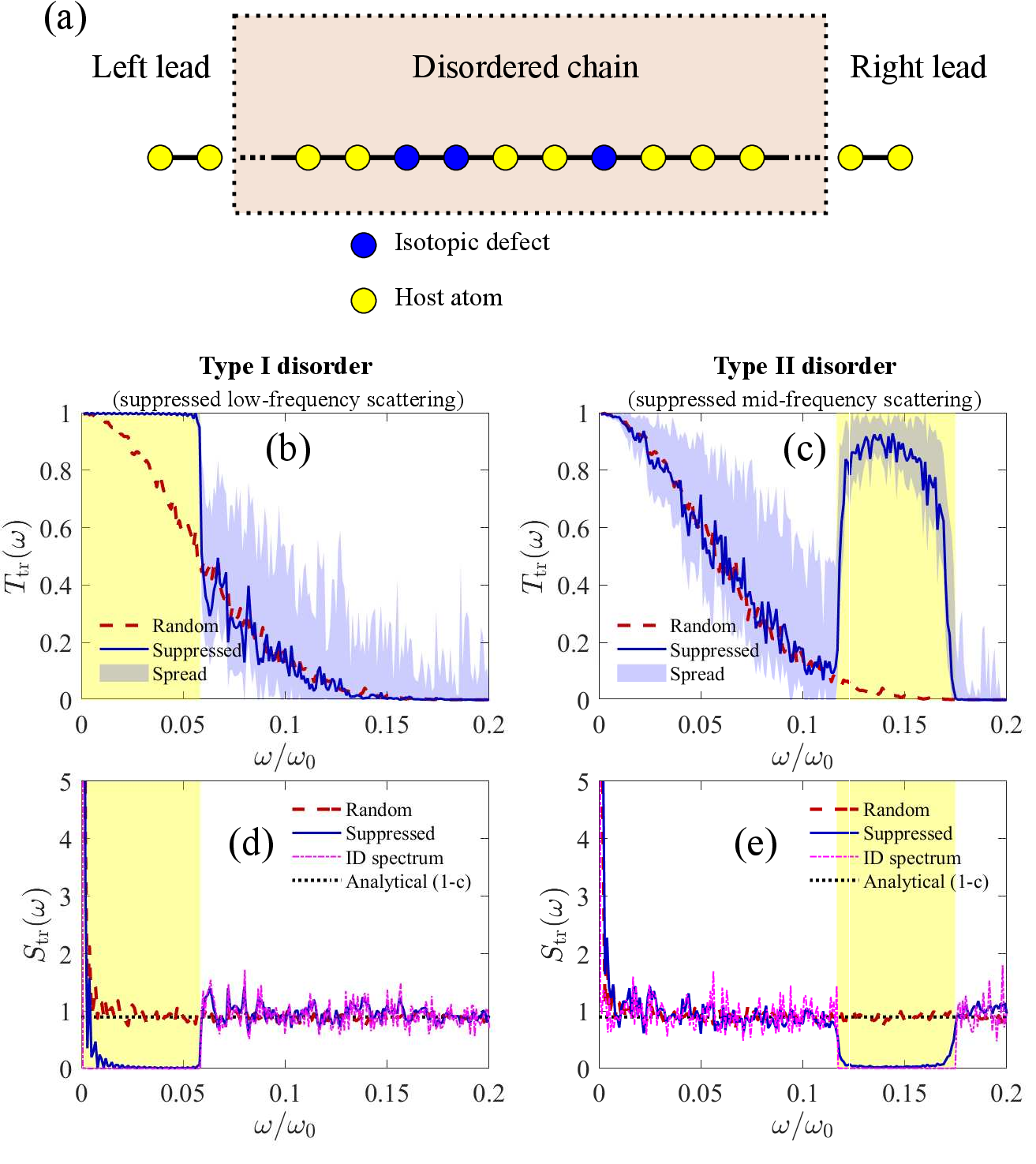}
\par\end{centering}
\caption{(a) Schematic of AGF simulation of the disordered chain sandwiched
between two leads. 
{
For (b) Type I and (c) Type II structured disorder,
we compare their log-averaged transmission spectra $T_{\text{tr}}(\omega)$
(solid line labeled ``Suppressed'') with $T_{\text{tr}}(\omega)$
for random disorder (dashed line labeled ``Random''), with spread indicated by the blue-shaded regions. 
}
The TFW for
suppressed scattering is shaded in yellow. We also compare the transmission-implied
structure factor $S_{\text{tr}}(\omega)$ for random disorder (dashed
line labeled ``Random'') to (d) Type I and (e) Type II disorder (solid
line labeled ``Suppressed''). We also plot the corresponding $\langle S_{\text{id}}(2k(\omega))\rangle$
(dot-dash line labeled ``ID spectrum'') 
{ 
and the thermodynamic limit for random disorder (dotted line labeled ``Analytical (1-c)'')}.
}

\label{fig:Results_SuppressedScattering}
\end{figure}

To validate our theory, we simulate
phonon transmission through a linear chain with ID's, using the Atomistic
Green's Function (AGF) method like in Refs.~\cite{WZhang:NHT07_Atomistic,ZYOng:JPCM14_Ballistic,ZYOng:PRB14_Enhancement}.
We set the concentration of IDs at $c=0.1$, similar to that in Ref.~\cite{ISavic:PRL08_CNTLocalization},
and the ratio of the isotopic defect mass to the mass of the host
atom as $m_{\text{id}}/m=2.6$, equal to that for germanium and silicon.
Figure~\ref{fig:Results_SuppressedScattering}(a) shows a schematic
of the disordered chain, which has 4000 atoms and a length of $L=4000$,
sandwiched between two semi-infinite pristine leads comprising the
host atoms. For each realization of structured disorder, we compute
the frequency-dependent transmission function $\Xi(\omega)$ over
the frequency range of $0<\omega\leq0.2\omega_{0}$ where $\omega_{0}=\sqrt{\zeta/m}$.
As a benchmark, we use the system where the IDs are randomly distributed. 

We simulate four types of structured disorder described in Table~\ref{tab:SimulationParameters}.
In each disorder type, we set the TFW to be between $\omega_{\text{lo}}$
and $\omega_{\text{hi}}$, which correspond to the wave numbers $k_{\text{lo}}=\frac{2}{a}\arcsin(\omega_{\text{lo}}/\omega_{\text{max}})$
and $k_{\text{hi}}=\frac{2}{a}\arcsin(\omega_{\text{hi}}/\omega_{\text{max}})$,
respectively. We define $k_{\text{lo}}$ and $k_{\text{hi}}$ with
respect to a characteristic wave number $k_{\text{cut}}=2\pi c/a$
determined by the ID concentration. 
{
We also generate an ensemble of 20 realizations for each disorder type, 
by minimizing Eq.~(\ref{eq:ObjectiveFunction}).
The values of the cost function before and after minimization are provided in Table~\ref{tab:OptimizationResults}, 
which shows that the final values of $E_\mathcal{K}$ are smaller for the low-frequency TFW, compared to the mid-frequency TFW. 
}

Type I and II disorder correspond
to how scattering can be suppressed for low-frequency and mid-frequency
modes in the TFW while Type III and IV correspond to enhanced low-frequency
and mid-frequency scattering. In particular, Type I disorder represents
a form of stealthy hyperuniform disorder~\cite{STorquato:PhysRep18_Hyperuniform,VRomero-Garcia:APLMater21_Wave}
for a 1-D lattice, with Type II disorder as a generalization for wave
numbers in the mid-frequency range. 

\begin{table}
\begin{tabular}{|c|c|c|c|}
\hline 
Type & $k_{\text{lo}}/k_{\text{cut}}$ & $k_{\text{hi}}/k_{\text{cut}}$ & Scattering\tabularnewline
\hline 
\hline 
I & $0$ & $0.05$ & Suppressed, low-frequency\tabularnewline
\hline 
II & $0.1$ & $0.15$ & Suppressed, mid-frequency\tabularnewline
\hline 
III & $0$ & $0.05$ & Enhanced, low-frequency\tabularnewline
\hline 
IV & $0.1$ & $0.15$ & Enhanced, mid-frequency\tabularnewline
\hline 
\end{tabular}

\caption{Description of structured disorder types used in our simulations.
The column for `Scattering' describes the effect of disorder on scattering
in the TFW. }
\label{tab:SimulationParameters}
\end{table}

{
\begin{table}
\begin{tabular}{|c|c|c|c|c|}
\hline 
\multirow{2}{*}{Type} & \multicolumn{2}{c|}{Initial $E_{\mathcal{K}}$} & \multicolumn{2}{c|}{Final $E_{\mathcal{K}}$}\tabularnewline
\cline{2-5} \cline{3-5} \cline{4-5} \cline{5-5} 
 & Mean & SD & Mean & SD\tabularnewline
\hline 
I & 1.71 & 0.191 & 0.00209 & 0.000722 \tabularnewline
\hline 
II & 1.73 & 0.148 & 0.00558 & 0.000508 \tabularnewline
\hline 
III & -0.00755 & 0.0879 & -11.8 & 0.485 \tabularnewline
\hline 
IV & -0.0382 & 0.101 & -8.19 & 0.166 \tabularnewline
\hline 
\end{tabular}

\caption{The mean and standard deviation (SD) for the initial and final values of cost function from Eq.~(\ref{eq:ObjectiveFunction}) over an ensemble of 20 initial configurations for each disorder type.}
\label{tab:OptimizationResults}
\end{table}
}

\section{Phononic transparency with structured disorder}
Figures~\ref{fig:Results_SuppressedScattering}(b) and (c) show the
log-averaged transmission spectra $T_{\text{tr}}(\omega)=\exp[\langle\log\Xi(\omega)\rangle]$~\cite{AAAsatryan:PRB11_Anderson,SGredeskul:LTP12_Anderson} 
for Type I (suppressed low-frequency scattering) and II (suppressed
mid-frequency scattering) disorder described in Table~\ref{tab:SimulationParameters}.
In Fig.~\ref{fig:Results_SuppressedScattering}(b), we observe near-total
phononic transparency ($T_{\text{tr}}\sim1$) within the TFW for Type
I disorder. Outside of the TFW, $T_{\text{tr}}(\omega)$ for the structured
disorder is comparable to $T_{\text{tr}}(\omega)$ for random disorder. 

For Type II disorder in Fig.~\ref{fig:Results_SuppressedScattering}(c),
the TFW is higher with a nontrivial lower frequency bound ($\omega_{\text{lo}}\neq0$).
In this case, $T_{\text{tr}}(\omega)$ in the TFW is not totally transparent
although it is exhibits much greater transmission enhancement compared
to $T_{\text{tr}}(\omega)$ for random disorder. This reduction in
phononic transparency is due to the greater difficulty in minimizing
Eq.~(\ref{eq:ObjectiveFunction}) for the mid-frequency wave numbers
in between $k_{\text{lo}}$ and $k_{\text{hi}}$ for Type II disorder.
The $\omega^{2}$ prefactor in Eq.~(\ref{eq:InvLocalLengthPowerSpectrum})
also also partially offsets the effect of the minimization of $S_{\text{id}}(2k)$.
In addition, we have tested our approach to scattering suppression
for TFWs at higher frequencies and found that the phononic transparency
diminishes at higher frequencies.

The frequency-dependent phononic transparency in Figs.~\ref{fig:Results_SuppressedScattering}(b)
and (c) is also reflected in the plot of the transmission-implied
structure function $S_{\text{tr}}(\omega)$, which is close to zero
and aligns closely with $\langle S_{\text{id}}(2k(\omega))\rangle$
in the TFW for $k(\omega)=\frac{2}{a}\arcsin(\omega/\omega_{\text{max}})$,
in Figs.~\ref{fig:Results_SuppressedScattering}(d) and (e). Outside
of the TFW, we find that $S_{\text{tr}}(\omega)$ fluctuates around
the value of $1-c$, which corresponds to the thermodynamic limit
of $S_{\text{id}}(2k(\omega))$ for random disorder. The results in
Fig.~\ref{fig:Results_SuppressedScattering} suggest that Eq.~(\ref{eq:InvLocalLengthPowerSpectrum})
reliably predicts phononic transparency.

\section{Phononic opacity with structured disorder} 
\begin{figure}
\begin{centering}
\includegraphics[width=8.6cm]{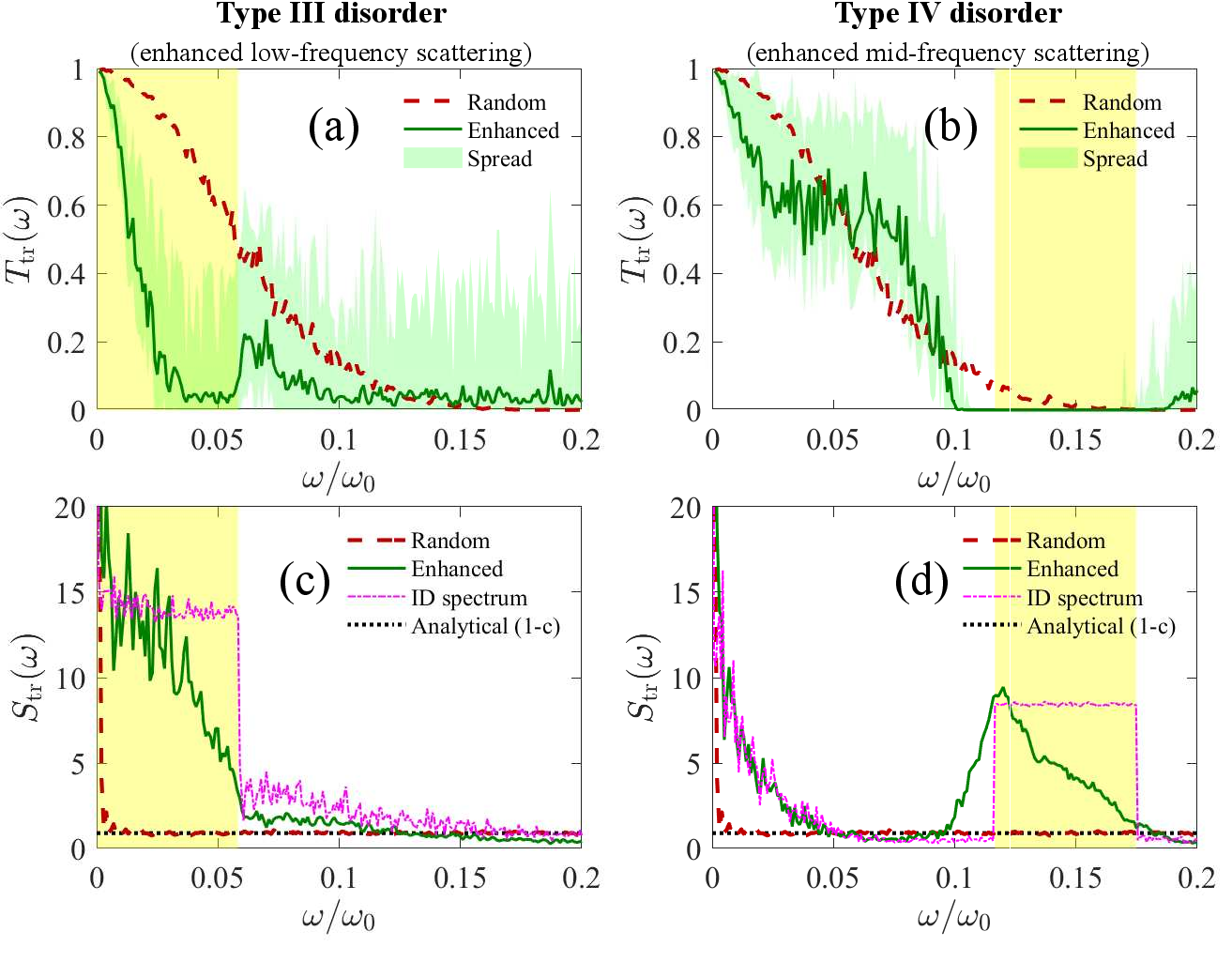}
\par\end{centering}
\caption{For (a) Type III and (b) Type IV structured disorder, we compare their
log-averaged transmission spectra $T_{\text{tr}}(\omega)$ (solid
line labeled ``Enhanced'') to $T_{\text{tr}}(\omega)$ for random
disorder (dashed line labeled ``Random''), 
{
with the spread in the former indicated by the green-shaded regions.
}
The TFW for enhanced scattering
is shaded in yellow. We also compare the transmission-implied structure
factor $S_{\text{tr}}(\omega)$ for random disorder (dashed line labeled
``Random'') to (c) Type III and (d) Type IV structured disorder (solid
line labeled ``Enhanced''). We also plot the corresponding $\langle S_{\text{id}}(2k(\omega))\rangle$
(dot-dash line labeled ``ID spectrum'') 
{
and the thermodynamic limit for random disorder (dotted line labeled ``Analytical (1-c)'').}
}

\label{fig:Results_EnhancedScattering}
\end{figure}

Figures~\ref{fig:Results_EnhancedScattering}(a)
and (b) show the log-averaged transmission spectra $T_{\text{tr}}(\omega)$
for Type III (enhanced low-frequency scattering) and IV (enhanced
mid-frequency scattering) disorder. Unlike the previous subsection,
the enhanced scattering impedes transmission and increases phononic
\emph{opacity} in the TFW.

In Fig.~\ref{fig:Results_EnhancedScattering}(a), we observe a substantial
reduction in $T_{\text{tr}}(\omega)$ for Type III disorder, compared
to $T_{\text{tr}}(\omega)$ for random disorder, within the TFW although
there is no total phononic opacity. Outside of the TFW, the $T_{\text{tr}}(\omega)$
is still significantly reduced compared to $T_{\text{tr}}(\omega)$
for random disorder, in contrast to the case for Type I disorder in
Fig.~\ref{fig:Results_SuppressedScattering}(b) where the $T_{\text{tr}}(\omega)$'s
for structured and random disorder are close. This suggests that even
though we have optimized the disorder for enhanced scattering in the
TFW, the disorder still generates enhanced scattering in the other
parts of the frequency spectrum. Figure~\ref{fig:Results_EnhancedScattering}(b)
shows the transmission spectrum for the TFW associated Type IV (enhanced
mid-frequency scattering) disorder. In this case, we find that transmission
($T_{\text{tr}}\sim0$) is completely blocked within the TFW.
{
The blockage of the phonon transmission in the TFW can also be characterized by its insertion loss spectra, 
which is shown in Figs. S1 and S2 in the Supplementary Material.
}

Figures~\ref{fig:Results_EnhancedScattering}(c) and (d) show a significant
discrepancy between $S_{\text{tr}}(\omega)$ and $\langle S_{\text{id}}(2k(\omega))\rangle$
within the TFW. In Fig.~\ref{fig:Results_EnhancedScattering}(c),
$\langle S_{\text{id}}(2k(\omega))\rangle$ is approximately constant
and higher than $S_{\text{tr}}(\omega)$ for random disorder within
the TFW. The difference is also observed for Type IV disorder in Fig.~\ref{fig:Results_EnhancedScattering}(d).
Unlike $\langle S_{\text{id}}(2k(\omega))\rangle$ in the TFW, $S_{\text{tr}}(\omega)$
is not constant and decreases with frequency. 
{
The discrepancy could be due to the inadequacy of Eq.~(\ref{eq:InvLocalLengthPowerSpectrum}), 
which assumes weak disorder and scattering.
This assumption is not valid for Type III and Type IV disorder, which enhance scattering within the TFW. 
}

\section{Phonon localization} 
To relate the phononic transparency
and opacity to the spatial localization of phonons, we characterize
the localization of the eigenmodes in the \emph{finite} disordered
chain segment from Fig.~\ref{fig:Results_SuppressedScattering}(a),
by computing the inverse participation ratio (IPR) for each eigenmode.
In the chain segment, the leftmost and rightmost atoms are coupled
to fixed walls, 
{
like in Ref.~\cite{CMonthus:PRB10_ParticipRatios}, 
} 
after the leads are detached. The normalized eigenmode
$u_{i}$ with eigenfrequency $\omega_{i}$ satisfies the eigenvalue
equation $\mathbf{K}u_{i}=\mathbf{M}\omega_{i}^{2}u_{i}$, where $\mathbf{K}$
and $\mathbf{M}$ denote the interatomic force constant and mass matrix,
respectively. For each $u_{i}$, we compute its IPR $\Phi_{i}=\sum_{n=1}^{N}|u_{i}^{n}|^{4}$,~\cite{CMonthus:PRB10_ParticipRatios}
where $u_{i}^{n}$ is the eigenvector component for atom $n$. If
the mode is fully delocalized (localized), then we expect $\Phi_{i}\sim1/N$
($\Phi_{i}\sim1$). 

\begin{figure}
\begin{centering}
\includegraphics[width=8cm]{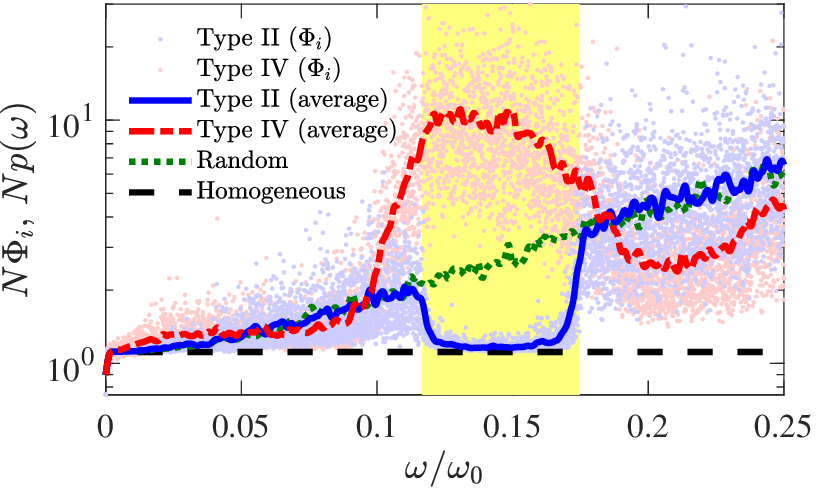}
\par\end{centering}
\caption{Comparison of the IPR $\Phi_{i}$ for Type II (blue dots) and IV (red
dots) disorder without and within the TFW (yellow-shaded region).
We also plot the average IPR $p(\omega)$ for homogeneous (black dashed
line), purely random (green dotted line), Type II (blue solid line)
and IV (red dash-dot line) disordered systems.}

\label{fig:Results_IPR}
\end{figure}

{
Figure~\ref{fig:Results_IPR} shows the scatter plot of $D = 80000$ data
points of $(\omega_{i},N\Phi_{i})$ from the ensembles for Type II
and IV disorder.
}
To see the trend more clearly for each disorder type,
we plot the average frequency-dependent IPR $N p(\omega)$, in
which we define $p(\omega)=\frac{1}{C(\omega)}\sum_{i=1}^{D}\Phi_{i}g_{i}(\omega)$
with $g_{i}(\omega)=\exp[-(\omega-\omega_{i})^{2}/\Delta^{2}]$ for
$\Delta=0.001\omega_{0}$ and $C(\omega)=\sum_{i=1}^{D}g_{i}(\omega)$.
We also plot $N p(\omega)$ for the purely random and homogeneous
systems as benchmarks. 
{
In addition, to characterize how the scattering affects the distribution of eigenfrequencies within the TFW,
we also plot the density of states in Fig.~S3 of the Supplementary Material. 
}

In general, random disorder leads to a monotonic increase of $p(\omega)$
with $\omega$. We observe that $p(\omega)$ for Type II disorder
agrees closely with $p(\omega)$ for purely random disorder outside
of the TFW. However, within the TFW, $p(\omega)$ for Type II disorder
agrees with the homogeneous system, i.e. $p\sim1/N$, indicating the
absence of localization and consistent with the phononic transparency
in Fig.~\ref{fig:Results_SuppressedScattering}(c). In contrast,
we observe an increase in $p(\omega)$ for Type IV disorder within
the TFW, indicating an enhancement in localization consistent with
the phononic opacity in Fig.~\ref{fig:Results_EnhancedScattering}(b).

\section{Conclusions} 
We have demonstrated an approach,
based on the static structure factor of the isotopic defect distribution,
for controlling phonon scattering in a targeted frequency
window to create phononic transparency or opacity. 
The enhanced (reduced) scattering also increases (decreases) the localization of phonon modes with eigenfrequencies in the TFW.
This approach may
be extended to control thermal and coherent phonon transport in aperiodic
superlattices,~\cite{YWang:PRB14_Decomposition,PRChowdhury:NE20_MachineLearning,RHu:PRX20_Machine,RHu:PRB21_Direct,TMaranets:APL24_Prominent}
by arranging the atomic layers in a similar manner to suppress or
enhance scattering, or to enhance thermoelectric performance by selectively
blocking phonon transmission.

\acknowledgements
{
This work was supported by Singapore NRF Thematic Competitive Research Programme (Grant No. T-CRP-2025-0030).
}


\bibliographystyle{eplbib}
\bibliography{PaperReferences}

\end{document}


\maketitle


\section{Insertion loss (IL) spectra}

\begin{figure}
\begin{centering}
\includegraphics[width=7.5cm]{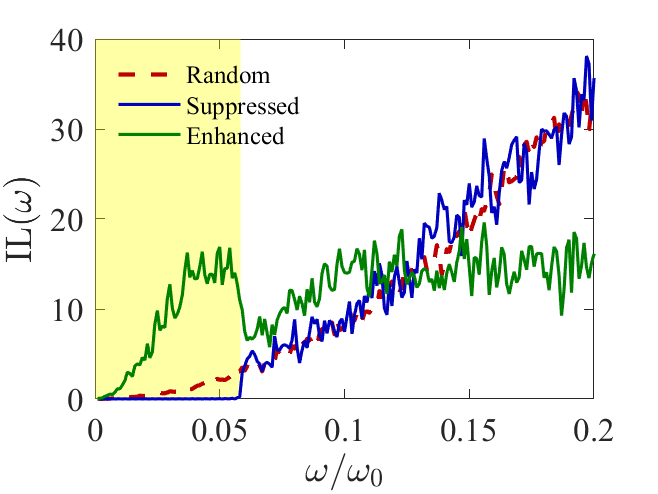}
\par\end{centering}
\caption{Insertion loss spectrum for purely random (dashed line labeled `Random'), Type I (solid blue line labeled `Suppressed') and Type III (solid green
line  labeled `Enhanced') disorder within and without the low-frequency TFW (yellow-shaded region). 
Type I and III disorder correspond to suppressed and enhanced scattering in the TFW, respectively.
}
\label{fig:IL_iso}
\end{figure}

\begin{figure}
\begin{centering}
\includegraphics[width=7.5cm]{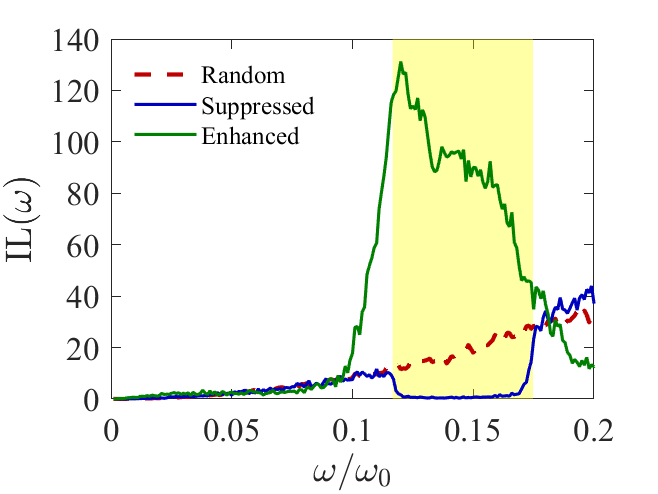}
\par\end{centering}
\caption{Insertion loss spectrum for purely random (dashed line labeled `Random'), Type II (solid blue line labeled `Suppressed') and Type IV (solid green
line  labeled `Enhanced') disorder within and without the mid-frequency TFW (yellow-shaded region). 
Type II and IV disorder correspond to suppressed and enhanced scattering in the TFW, respectively.
}

\label{fig:IL_poc2}
\end{figure}

In Figs.~\ref{fig:IL_iso} and \ref{fig:IL_poc2}, we plot the spectra of the frequency-dependent  IL
which characterizes the degree of attenuation by the disorder and is defined as 
\begin{equation}
    \text{IL}(\omega) = -10 \log_{10} T_\text{tr} (\omega) \ ,
\label{eq:ILDefine}
\end{equation}
where $T_\text{tr} (\omega)$ denotes the log-averaged transmission function, for the different disorder types. 
The IL spectra complement the transmission $T_\text{tr}(\omega)$ spectra in Figs. 1 and 2 of the paper
and quantifies the degree of attenuation which may be difficult to characterize when $T_\text{tr}(\omega) \ll 1$ in the targeted frequency window (TFW) where scattering is enhanced.

Figure~\ref{fig:IL_iso} shows the average IL spectra for purely random, Type I (suppressed low-frequency scattering) and Type III (enhanced low-frequency scattering) disorder.
Figure~\ref{fig:IL_poc2} shows the IL spectra for purely random, Type II (suppressed mid-frequency scattering) and Type IV (enhanced mid-frequency scattering) disorder.
We observe that within the TFW in Fig.~\ref{fig:IL_poc2} for Type IV disorder, the IL spectrum decreases monotonically with frequency, 
consistent with the observation for the $S_\text{tr}(\omega)$ spectrum for Type IV disorder in Fig.~2(d) of the paper.

\section{Mean density of states}

For each type of disorder (I to IV), we define the normalized mean density of states (DOS) as 
\begin{equation}
    G(\omega) = \frac{C(\omega)}{\int_0^\infty C(\omega) d\omega} \ ,
\label{eq:MeanDOS}
\end{equation}
where 
\begin{equation}
    C(\omega) = \sum_{i=1}^{D} \exp \left[ -\frac{(\omega-\omega_i)^2}{\Delta^2} \right] 
\label{eq:SumModalDOS}    
\end{equation}
for $\Delta = 0.001 \omega_0$. 
In Eq.~(\ref{eq:SumModalDOS}),
the variable $D$ denotes the total number of modes in the ensemble for each disorder type 
and has a numerical value of $D = 80000$ as we use 20 realizations of an $N = 4000$ system.
The variable $\omega_i$ denotes the eigenfrequency of the $i$-th mode of the ensemble.

\begin{figure}
\begin{centering}
\includegraphics[width=8cm]{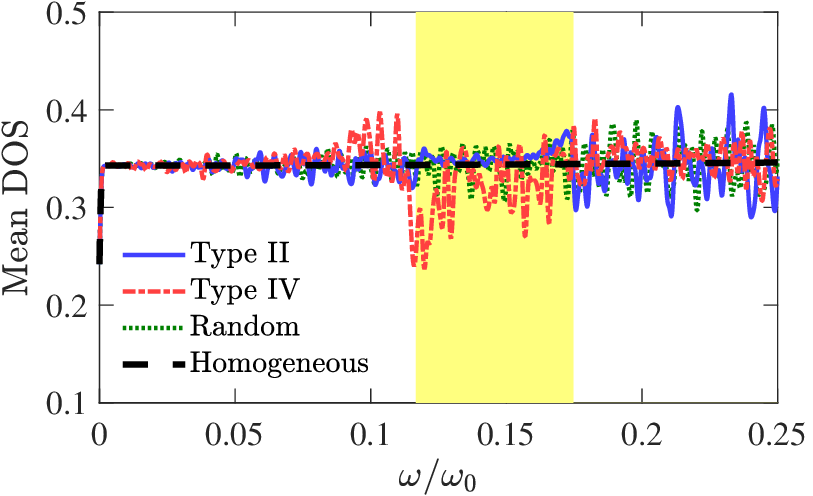}
\par\end{centering}
\caption{Comparison of the normalized mean DOS for the homogeneous (black dashed
line), purely random (green dotted line), Type II (blue solid line)
and IV (red dash-dot line) disordered systems within and without the mid-frequency TFW (yellow-shaded region).
}
\label{fig:Mean_DOS}
\end{figure}

To complement Fig. 3 in the paper, 
we plot the spectrum of $G(\omega)$ from Eq.~(\ref{eq:MeanDOS}) in Fig.~\ref{fig:Mean_DOS} 
for Type II (suppressed mid-frequency scattering) and IV (enhanced mid-frequency scattering) disorder
and compare them to the spectra for a homogeneous system and a purely random system in the low-frequency range.

As expected, the $G(\omega)$ for the homogenous system is effectively a constant in the low-frequency range. 
And likewise, so is the $G(\omega)$ for the purely random system although it fluctuates around the $G(\omega)$ for the homogenous system.
For the Type II and IV disordered systems, the $G(\omega)$ is also effectively a constant as $\omega \rightarrow 0$.

Within the TFW, the $G(\omega)$ for the Type II disordered system is close to the one for the homogenous system, 
possibly because of the suppressed scattering in for the modes in the TFW.
On the other hand, we observe a noticeable reduction in the $G(\omega)$ for the Type IV disordered system, 
especially when $\omega$ is close to $\omega_\text{lo}$, the lower bound of the TFW.
Tentatively, we attribute this reduction in $G(\omega)$ to the enhanced scattering in the TFW, 
which results in greater shifts in the eigenfrequencies. 
A rigorous explanation of this phenomenon is left for future work.

